\documentclass[english,12pt,a4paper]{article}
\usepackage{authblk}
\usepackage[text={17.0cm,23.7cm}]{geometry}
\usepackage[T1]{fontenc}
\usepackage[utf8]{inputenc}
\usepackage[english]{babel} 
\usepackage{hyperref}
\usepackage{graphicx}
\usepackage{amsmath,amssymb}
\usepackage[dvipsnames]{xcolor}
\usepackage{yfonts}
\usepackage{ulem}
\usepackage{enumitem}
\usepackage[mathscr]{euscript}
\usepackage{float}
\usepackage{csquotes}
\usepackage{cleveref} 
\usepackage[nottoc]{tocbibind}

\begin{document}
	
	\title{\vspace{-2cm}
		\vspace{0.6cm}
	\textbf{ Attenuation of the ultra-high-energy neutrino flux by dark matter scatterings}\\[8mm]}
\author[1,2]{Ivan Esteban}
\author[3]{Alejandro Ibarra}

\affil[1]{\normalsize\textit{Department of Physics, University of the Basque Country UPV/EHU, PO Box 644, 48080 Bilbao, Spain}}
\affil[2]{\normalsize\textit{EHU Quantum Center, University of the Basque Country UPV/EHU}}
\affil[3]{\normalsize\textit{Physik-Department, Technische Universit\"at M\"unchen, James-Franck-Stra\ss{}e, 85748 Garching, Germany}}

\date{}
\maketitle

\begin{abstract}
A flux of ultra-high-energy (UHE) neutrinos, produced by astrophysical sources at cosmological distances, is anticipated to exist and reach Earth. In this paper, we investigate the impact on the total flux, energy spectrum, and arrival directions of UHE neutrinos of neutrino-dark matter (DM) scatterings. We study scatterings both in the intergalactic medium and in the Milky Way. We emphasize the complementarity among neutrino detectors at different latitudes, that can probe anisotropies induced by neutrinos scattering with the Milky Way DM halo. We also discuss that, with mild astrophysical assumptions, limits on the DM-$\nu$ scattering cross section can be placed even if the neutrino sources are unknown. Finally, we explore all this phenomenology with the recent UHE neutrino event KM3230213A, and place the corresponding limits on the DM-$\nu$ scattering cross section.
\end{abstract}

\section{Introduction}
\label{sec:Intro}

The origin of the ultra-high-energy cosmic rays (UHECRs) remains a mystery to this day. The most accepted explanation is that UHECRs are accelerated in extreme astrophysical environments via electromagnetic processes like Fermi acceleration (for reviews, see, {\it e.g.}, Refs.~\cite{Kotera:2011cp,Anchordoqui:2018qom}). If this is the case, UHECRs are expected to interact with the radiation and matter surrounding the sources, as well as with background radiation en-route to Earth, producing a flux of ultra-high energy (UHE) neutrinos that could be detected at Earth. 

In the Standard Model of Particle Physics neutrinos interact very weakly with matter and radiation, and the UHE neutrino flux is expected to arrive at Earth isotropically and with negligible attenuation. However, it is now firmly established that the Universe is permeated by an additional component of matter, called Dark Matter (DM). While the interactions of neutrinos with ordinary matter are well-described by electroweak theory (at least up to the energies that have been probed), the interactions of neutrinos with DM are completely unknown. Upper limits on the DM-neutrino cross section have been derived for neutrino energies ${E_\nu\sim 100\,\mathrm{eV}}$ from the  Lyman-alpha forest \cite{Wilkinson:2014ksa} and from the number of Milky Way dwarf spheroidal galaxies \cite{Akita:2023yga, Crumrine:2024sdn}; for $E_\nu \sim 10$ MeV from supernova neutrinos \cite{Raffelt:1996wa,Heston:2024ljf}; for $E_\nu\sim 1$ TeV from the Seyfert Galaxy NGC 1068~\cite{Cline:2023tkp} and for $E_\nu\sim 100$ TeV from the blazar TXS 0506+056 \cite{Ferrer:2022kei,Cline:2022qld}. However, the scattering cross section for $E_\nu\gtrsim 100$ TeV remains totally unconstrainted, save for extrapolations of other constraints to higher energies assuming a specific energy dependence of the cross section. 

In this paper, we investigate the attenuation of the UHE neutrino flux due to scatterings with DM in both the intergalactic medium and the Milky Way halo. Our analysis is conservative, as we do not include potential further attenuation effects near the (so far unknown) sources of UHE neutrinos. We argue that DM–neutrino interactions can lead to significant and unique modifications to the observed flux, energy spectrum, and arrival directions of UHE neutrinos. As a case study, we focus on the recently reported event KM3-230213A~\cite{KM3NeT:2025npi}, including its consistency with null results from IceCube and the Pierre Auger Observatory~\cite{IceCubeCollaborationSS:2025jbi,PierreAuger:2019azx}. We show that, with mild astrophysical assumptions, this observation sets limits on DM-neutrino scattering even if its origin is unknown.

The paper is organized as follows. In Section \ref{sec:attenuation_general} we analyze the attenuation of a neutrino flux generated at high redshift due to scatterings with DM in the intergalactic medium and in the Milky Way. In Section \ref{sec:attenuation_UHE} we apply this formalism to investigate the implications of the attenuation of a diffuse UHE flux, arguing that the Waxman-Bahcall bound allows to set limits on neutrino-DM scattering even if the unattenuated flux is unknown. We apply this idea to a concrete scenario in which the event KM3-230213A could be part of a diffuse flux, setting the corresponding limit. In Section \ref{sec:complementarity} we emphasize the importance of a network of neutrino detectors at different latitudes to robustly establish whether the UHE neutrino flux is affected by attenuation due to DM particles. Finally, in Section \ref{sec:conclusions}, we present our conclusions. 

\section{Neutrino flux attenuation by dark matter scatterings}
\label{sec:attenuation_general}

Let us assume that neutrinos are injected into the Universe at cosmological time $t$ with energy $E$, with comoving rate ${\cal I}(E,t)$. The time evolution of the comoving number density of neutrinos is governed by the following transport equation~\cite{Berezinsky:2005fa}
\begin{equation}
  \frac{\partial n_\nu(E, t)}{\partial t} = \mathcal{I}(E, t) + \frac{\partial}{\partial E} \left[ H(t) E n_\nu(t, E) \right] - n_\mathrm{DM}(t) \sigma(E) n_\nu(E, t) \, ,
\end{equation}
where we have included the neutrino energy redshift due to the expansion of the Universe, with rate $H(t)$; and the decrease in the neutrino number density due to interactions with DM particles with density $n_{\rm DM}(t)$ and energy-dependent cross section $\sigma(E)$. Here we neglect for simplicity the production of a secondary neutrino flux at lower energies in the scattering process. Since the astrophysical neutrino flux is expected to fall with energy and the full transport equation for elastic scattering conserves particle number, this secondary flux is expected to be small with respect to the more abundant lower-energy primary flux. \footnote{As an explicit quantitative example, if the primordial neutrino flux falls as $E^{-2}$, the amount of neutrinos with energy higher than 2.6\,EeV (i.e., above the energy of the KM3NeT event) is only 3\% of the amount of neutrinos with energies between 0.072 and 2.6\,EeV (i.e., the energy range of the KM3NeT event).}

The transport equation can be rewritten in terms of redshift as
\begin{equation}
  -H(z) (1+z) \frac{\partial n_\nu(E, z)}{\partial z} = \frac{\partial}{\partial E} \left[ H(t) E n_\nu(E, z) \right] + \mathcal{I}(E, z) - n_\mathrm{DM}(z) \sigma(E) n_\nu(E, z) \, ,
\end{equation}
which has as solution
\begin{equation}
  n_\nu(E, z) = \frac{1}{1+z} \int_z^{z_{\rm max}} \frac{\mathrm{d}z'}{H(z')} \mathcal{I}\left(\frac{1+z'}{1+z}E, z'\right) \exp\left[-\displaystyle \int_z^{z'} \frac{\mathrm{d}z''}{H(z'')(1+z'')} n_\mathrm{DM}(z'') \sigma\left(\frac{1+z''}{1+z}E\right)\right]\,,
\end{equation}
where we have assumed that neutrinos are injected only at redshifts $z\leq z_{\rm max}$. This solution can be readily obtained by the change of variables $\mathcal{Z}(E, z) \equiv (1+z) n\big(E(1+z), z\big)$~\cite{Ahlers:2009rf}. The prefactor $(1+z')/(1+z)$ multiplying the neutrino energy $E$ corresponds to energy redshift due to the expansion of the Universe, and the exponential term captures neutrino attenuation due to scattering with DM particles.

The differential neutrino flux at a terrestrial detector (located at $z=0$) can be straightforwardly calculated using that neutrinos move relativistically, so that $d\phi/dE=n_\nu(E,0)$:
\begin{align}
    \frac{\mathrm{d}\phi}{\mathrm{d}E}(E,z=0) & = \int_0^{z_{\rm max}} \frac{\mathrm{d}z'}{H(z')} \mathcal{I}\big((1+z')E, z'\big) \exp\left[-\displaystyle \int_0^{z'} \frac{\mathrm{d}z''}{H(z'')(1+z'')} n_\mathrm{DM}(z'') \sigma\big((1+z'')E\big)\right] \nonumber \\
    & \equiv \int_0^{z_{\rm max}} \frac{\mathrm{d}z}{H(z)} \mathcal{I}\big((1+z)E, z\big) e^{-\tau(E, z)} \, .
  \label{eq:flux_solved}
\end{align}
Here, we have introduced for convenience an energy-dependent optical depth
\begin{align}
    \tau(E, z) & \equiv \int_0^{z} \frac{\mathrm{d}z'}{H(z')(1+z')} n_\mathrm{DM}(z') \sigma\big((1+z')E\big) \nonumber\\
    & = \frac{\rho_{\mathrm{DM},\, 0}}{M_\mathrm{DM} H_0}\int_0^{z} \frac{(1+z')^2 \sigma\big((1+z')E\big)}{\sqrt{\Omega_{\Lambda,0} + \Omega_{m,0}(1+z')^3}} \,\mathrm{d}z'\, ,
\end{align}
where we have expressed the DM number density in terms of the mass density $n_{\rm DM}(z)=\rho_{\rm DM}(z)/M_{\rm DM}$, and we have used that the mass density and the expansion rate scale with redshift as $\rho_{\rm DM}(z)=\rho_{{\rm DM},0} (1+z)^3$ and $H(z)=H_0\sqrt{\Omega_{\Lambda,0}+\Omega_{\rm m,0}(1+z)^{3}}$. $H_0$, $\Omega_{\Lambda,0}$, $\Omega_{{\rm m},0}$ and $\rho_{{\rm DM},0}$ are, respectively, the Hubble constant, the density parameters of cosmological constant and matter, and the mass density of DM; all at the present cosmic time. In our numerical analysis, we use $H_0=67.7\,\mathrm{km/s/Mpc}$, $\Omega_{\Lambda,0}=0.689$, $\Omega_{{\rm m},0}=0.311$, and $\rho_{{\rm DM},0}=1.26\times 10^{-6}\,\mathrm{GeV/cm^3}$~\cite{Planck:2018vyg}.

If the neutrino-DM cross section has a simple power-law dependence with neutrino energy, $\sigma_n(E) = \sigma(E_0) \left(\frac{E}{E_0}\right)^n$, one obtains
\begin{equation}
    \tau(E, z)  =  \frac{\sigma(E)}{M_\mathrm{DM}} \ell_n(z)\;,
    \label{eq:tau_z}
\end{equation}
where 
\begin{align}
\ell_n(z) \equiv \frac{\rho_{\mathrm{DM},\, 0}}{H_0}\int_0^{z} \frac{(1+z')^{2+n}}{\sqrt{\Omega_\Lambda + \Omega_m(1+z')^3}} \,\mathrm{d}z'
\end{align}
can be interpreted as an effective column density traversed by a neutrino produced at redshift $z$ when the cross section is proportional to $E^n$. The factor $(1+z')^n$ captures the fact that neutrino energy, and hence the neutrino-DM cross section, depends on redshift.

Once neutrinos enter the Milky Way, their flux is subject to an additional attenuation due to scatterings with DM particles in the halo of our galaxy. 
Since the Earth is not located at the center of the Milky Way, the attenuation depends on the arrival direction. Using that the flux at the halo boundary (which we identify with the virial radius $R_{\rm vir}\simeq 198$ kpc~\cite{Karukes:2019jwa}) is isotropic and given by Eq.~(\ref{eq:flux_solved}), the flux at Earth can be calculated from
\begin{align}
\frac{d\phi_{\rm Earth}}{dE}(\Omega,E)=
\frac{d\phi}{dE}(E,z=0) e^{-\tau_{\rm MW}(\Omega,E)}\;,
\label{eq:flux_attenuation_MW}
\end{align}
where the optical depth due to scattering with the DM in the Milky Way can be cast in an analogous way to Eq.~(\ref{eq:tau_z}),
\begin{equation}
    \tau_\mathrm{MW}(\Omega, E) =  
    \frac{\sigma(E)}{M_\mathrm{DM}} \ell(\alpha,\beta)\;.
\end{equation}
Here $\ell(\alpha,\delta)$ is the DM column density traversed by the neutrino flux in the Milky Way in the direction with right ascension $\alpha$ and declination $\delta$ as seen from Earth,
\begin{align}
\ell(E,\alpha,\delta)\equiv\int_0^{R_{\rm vir}}  \rho_{\mathrm{DM}}\big({\vec r}(s,\alpha,\delta)\big) \, \mathrm{d}s\, ,
\end{align} 
where $\rho_\mathrm{DM}(\vec r)$ is the Milky Way DM mass density and $\vec{r}(s)$ is expressed in Cartesian Galactic coordinates as $\vec{r}(s, \alpha, \delta)=s\,(\cos\delta\cos\alpha,\cos\delta\sin\alpha, \sin\delta) + \vec{r}_\odot$. $s$ is the distance along the line of sight, and $\vec{r}_\odot = -r_\odot\,(\cos \delta_\mathrm{GC} \cos\alpha_\mathrm{GC}, \cos\delta_\mathrm{GC} \sin \alpha_\mathrm{GC}, \sin \delta_\mathrm{GC})$ is the Earth position, with $r_\odot=8.18\,\mathrm{kpc}$, $\delta_\mathrm{GC} = -28.92^\circ$, and $\alpha_\mathrm{GC} = 266.42^\circ$ the coordinates of the Galactic Center as seen from Earth~\cite{Gravity:2019nxk}. In our analysis we adopt two possible choices for the DM density profile in the Milky Way, corresponding to a cuspy profile and to a cored profile. For the cuspy profile, we adopt the Navarro-Frenk-White (NFW) profile~\cite{Navarro:1995iw,Navarro:1996gj},
\begin{align}
\rho(\vec{r}) = \frac{\rho_0}{\left(\dfrac{r}{r_s} \right) \left( 1 + \dfrac{r}{r_s}\right)^2}\;,
\end{align}
with density parameter $\rho_0=1.15\times10^7\,M_\odot/\mathrm{kpc}$ and scale radius $r_s=14.8\,\mathrm{kpc}$~\cite{Cautun:2019eaf}; while for the cored profile we adopt the Burkert profile~\cite{Burkert:1995yz},
\begin{align}
\rho(\vec{r}) = \frac{\rho_0}{\left(1 + \dfrac{r}{r_c}\right)\left[1 + \left( \dfrac{r}{r_c} \right)^2\right]}\;,
\end{align}
with $\rho_0=4.8\times10^7\,M_\odot/\mathrm{kpc^3}$ and $r_c$ = 8.24 kpc \cite{Cautun:2019eaf, Karukes:2019jwa}.

\begin{figure}[t!]
    \centering
    \includegraphics[width=0.5\linewidth]{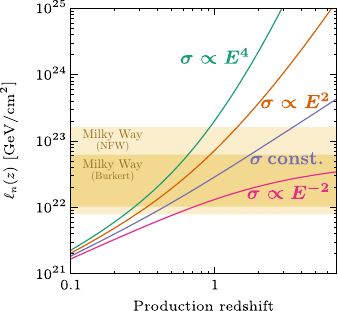}
    \caption{Effective integrated DM column density of a high-redshift source, for different energy-dependent cross sections. The yellow bands indicate the ranges of the DM column density in the Milky Way for an NFW or a Burkert profile, depending on the neutrino arrival direction.}
    \label{fig:column-density}
\end{figure}

We show in Fig.~\ref{fig:column-density} the effective column density $\ell_n(z)$ traversed by neutrinos produced at redshift $z$ for energy dependences of the scattering cross section, $n=4,2,0,-2$. For comparison, we also show in yellow the column density traversed in the Milky Way, where the different values in the shaded area correspond to different neutrino arrival directions. As apparent from the plot, the effective column density in the intergalactic medium is subdominant (albeit non-negligible) if neutrinos are produced at redshift $z\lesssim 0.2$, while it can be dominant if neutrinos are produced at redshift $z\gtrsim 1$, especially when the scattering cross section has a strong dependence with energy. The effective column density is in the range $10^{22}-10^{25}\,{\rm GeV}/{\rm cm}^2$. Therefore, for DM masses $M_{\rm DM}= 1$ GeV, the attenuation can be sizable for scattering cross sections larger than the Thomson cross section $\sigma_{\rm T}=6.6\times 10^{-25}\,{\rm cm}^2$. We present a detailed discussion of phenomenologically relevant cross sections in Section \ref{sec:attenuation_UHE}.

\begin{figure}[t!]
    \centering
    \includegraphics[width=0.49\linewidth]{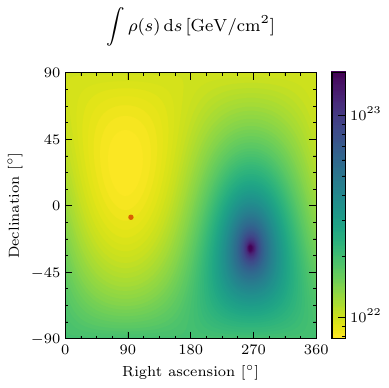} \includegraphics[width=0.49\linewidth]{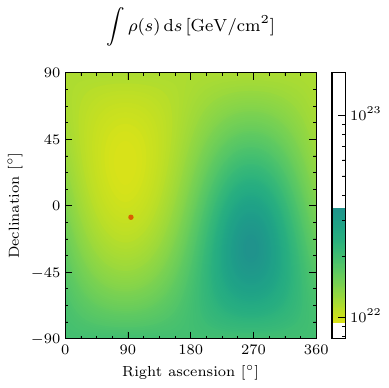}
    \caption{Integrated Milky Way DM column density at different directions, for an NFW (left) and Burkert (right) profile.  The red dot corresponds to the arrival direction of KM3‑230213A.}
    \label{fig:MW-map}
\end{figure}

The column density in the Milky Way DM halo depends on the arrival direction, which is shown in Fig. \ref{fig:MW-map} for the NFW profile (left plot) and the Burkert profile (right plot). The column density is largest in the direction of the Galactic Center (with equatorial coordinates $\alpha\simeq 17^h 45^m \simeq 266^\circ$ and  $\delta\simeq -29^\circ$), and smallest in the opposite direction. The anisotropy in the column density is stronger for a cuspy profile than for a cored profile, although a cored profile has on average a larger column density because both profiles have the same total DM mass. We also include for reference the location of the highest-energy neutrino observed to date, KM3‑230213A, with arrival direction centered in $\alpha\simeq 94.3^\circ$ and $\delta\simeq -7.8^\circ$, and a 99\% containment region of $3^\circ$ radius~\cite{KM3NeT:2025bxl}.

\begin{figure}[t!]
    \centering
    \includegraphics[width=0.5\linewidth]{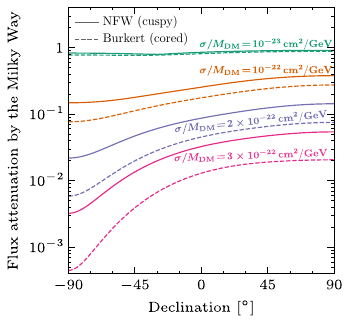}
    \caption{Neutrino-flux attenuation due to DM-neutrino scatterings in the Milky Way halo as a function of the declination, for an NFW profile (solid line) or a Burkert profile (dashed line),     averaged over one day (or, equivalently, averaged over right ascension).}
    \label{fig:attenuation_vs_declination}
\end{figure}

If the neutrino flux is isotropic at the halo boundary, the attenuation due to DM scattering would generate an anisotropic flux at the detector, {\it cf.} Eq.~(\ref{eq:flux_attenuation_MW}). In Fig. \ref{fig:attenuation_vs_declination} we show the flux attenuation as a function of the declination, integrating over a day (or, equivalently, integrating over the right ascension), for different values of the cross section per DM mass $\sigma/M_\mathrm{DM}$. For small values of $\sigma/M_{\rm DM}$, the attenuation due to the Milky Way is negligible and the flux at Earth is roughly identical to the flux at the halo boundary. In turn, as the cross section increases, there is a reduction of the flux from the North Celestial Pole ($\delta=90^\circ$); and an even larger reduction of the flux from the South Celestial Pole ($\delta=-90^\circ$), as an incoming neutrino from the latter direction traverses a larger fraction of the Milky Way DM halo, see Fig.~\ref{fig:MW-map}. Moreover, the anisotropy in the flux between these two directions becomes larger as the cross section increases. The impact is higher for a cored Burkert profile, which has an overall larger DM density than a cuspy NFW profile as mentioned above. For large cross sections, attenuation is exponentially sensitive to the DM density, further enhancing the difference among both profiles.

Thus, the number of events at a given detector that is more sensitive to a part of the sky depends critically on its location. This feature, that has been exploited within a single detector using high-energy astrophysical neutrino data~\cite{IceCube:2022clp, Arguelles:2017atb}, can be used to robustly probe if the UHE neutrino flux is attenuated by the Milky Way DM. We discuss this in detail in Section \ref{sec:complementarity}.

\section{Attenuation of the ultra-high energy neutrino flux}
\label{sec:attenuation_UHE}

In principle, using the observed UHE neutrino flux to explore the neutrino-DM cross section requires \textit{a priori} assumptions on the flux before attenuation. Here, we discuss how mild astrophysical assumptions allow to set conservative bounds.

Ultra-high energy neutrinos could be produced at astrophysical sources, when protons get accelerated to extremely high energies and collide with photons surrounding the source, thereby producing charged pions that eventually decay into neutrinos. Assuming that the sources are optically thin to protons, and in view to the measured flux of ultra-high energy cosmic rays, it is possible to derive a conservative upper bound on the expected ultra-high energy neutrino flux at Earth
\begin{align}
E_\nu^2 \, \Phi_\nu \lesssim 4.5 \times 10^{-8} \, \text{GeV}  \text{cm}^{-2}  \text{s}^{-1}  \text{sr}^{-1}\;;
\end{align}
this is the renowned Waxman-Bahcall bound \cite{Waxman:1998yy}. The actual flux is likely smaller, as the bound assumes strong redshift evolution and a proton-dominated composition of cosmic rays, among other conservative assumptions~\cite{Bahcall:1999yr}. The observation of a UHE neutrino flux could then be translated into limits on the DM-$\nu$ interaction cross section as we discuss below.

\begin{figure}[t!]
    \centering
    \includegraphics[width=0.49\linewidth]{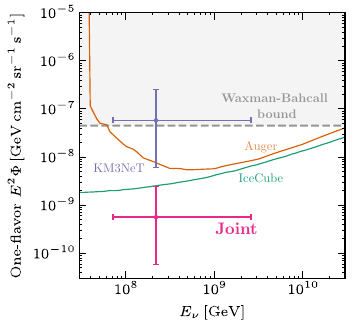}     \includegraphics[width=0.49\linewidth]{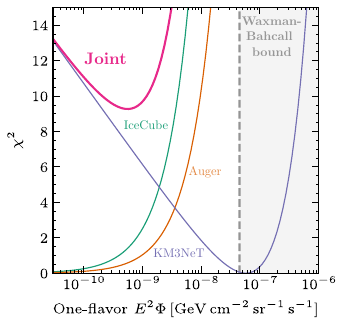}
    \caption{{\it Left panel:} Current landscape of UHE astrophysical neutrino fluxes, together with the Waxman-Bahcall upper bound. All limits and measurements are at 90\% CL. {\it Right panel:} $\chi^2$ of each experiment as well as the joint $\chi^2$ from KM3NeT, IceCube and the Pierre Auger Observatory.}
    \label{fig:landscape}
\end{figure}

The left panel of Fig.~\ref{fig:landscape} summarizes the landscape of current searches for an UHE astrophysical neutrino flux. The plots shows the latest 90\% C.L. upper limits from IceCube and the Pierre Auger Observatory, obtained following Ref.~\cite{Anchordoqui:2002vb} using the publicly available all-sky effective areas from Refs.~\cite{PierreAuger:2019azx, IceCubeCollaborationSS:2025jbi, KM3NeT:2025npi} as described in Appendix~\ref{app:Aeff}. We also show for comparison the Waxman-Bahcall bound. It is notable that current experiments are already sensitive to UHE neutrinos from astrophysical sources.

The figure also includes the neutrino event KM3-230213A, which is compatible with the Waxman-Bahcall bound, but it is however at odds with the upper limits from IceCube and the Pierre Auger Observatory. This tension is quantified in the right-plot, which shows a Poissonian $\chi^2$ for the number of events, assuming that the flux is proportional to $E^{-2}$ and integrating over the 90\% CL energy limits of the KM3NeT event, [0.072,2.60] EeV. As apparent from the Figure, the best-fit point from combining the observation of one event at KM3NeT and the non-observation of events at IceCube and the Pierre Auger Observatory hints towards much lower neutrino fluxes, approximately $\sim3\sigma$ smaller than the one reported by KM3NeT~\cite{Li:2025tqf, KM3NeT:2025ccp, IceCubeCollaborationSS:2025jbi}. We obtain the errors on the joint fit following the Feldman \& Cousins procedure \cite{Feldman:1997qc}, analogously to Ref.~\cite{KM3NeT:2025ccp} (we have checked that if we use the same IceCube dataset as Ref.~\cite{KM3NeT:2025ccp} we reproduce their results). Next-generation instruments will improve the IceCube sensitivity by two orders of magnitude \cite{Ackermann:2022rqc}, and will elucidate whether the UHE neutrino flux saturates the Waxman-Bahcall bound, as suggested by the observation of KM3-230213A.

\begin{figure}[t!]
    \centering
    \includegraphics[width=0.5\linewidth]{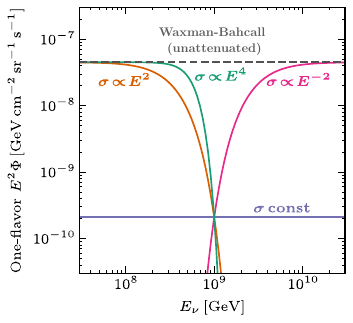}
    \caption{Sky-averaged neutrino-flux, attenuated by the Milky Way DM, for an unattenuated flux saturating the Waxman-Bahcall bound. We set $\sigma/M_\mathrm{DM} = 5 \times 10^{-22}\,\mathrm{cm^2/GeV} \times \left(\frac{E_\nu}{10^9\,\mathrm{GeV}}\right)^n$.}
    \label{eq:WB-attenuated}
\end{figure}

Regardless of the true nature of KM3-230213A, current instruments already provide valuable information on the possible accelerating mechanisms at the sources, as well as on the medium through which neutrinos propagate, and specifically the DM in the intergalactic medium and in the Milky Way.  In particular, if the unattenuated flux satisfies the Waxman-Bahcall bound, in the presence of sizable DM-$\nu$ interactions the bound on the attenuated flux at Earth will be more stringent. We show this in Fig.~\ref{eq:WB-attenuated} for the specific case of $\sigma/M_\mathrm{DM} = 5 \times 10^{-22}\,\mathrm{cm^2/GeV} \times (E_\nu/10^9\,\mathrm{GeV})^n$ and $n=-2,0,2,4$; where we have conservatively included only attenuation due to the Milky Way. Conversely, the detection of a UHE neutrino flux would automatically set an upper limit on $\sigma/M_\mathrm{DM}$, which would depend on the energy dependence of the cross section. 

\begin{figure}[hbtp]
    \centering
    \includegraphics[width=0.5\linewidth]{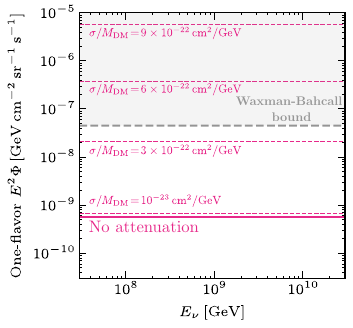}
    \caption{Unatennuated flux for different energy-independent DM-neutrino interaction cross sections, taking as a benchmark the current joint best-fit flux.}
    \label{fig:attenuation-KM3-event}
\end{figure}

\begin{figure}[hbtp]
    \centering
    \includegraphics[width=0.49\linewidth]{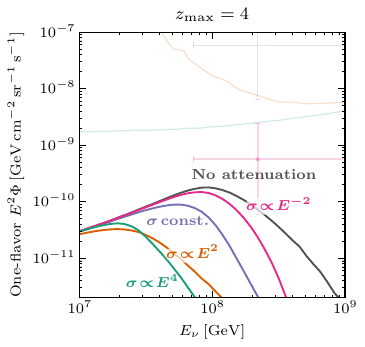} \includegraphics[width=0.49\linewidth]{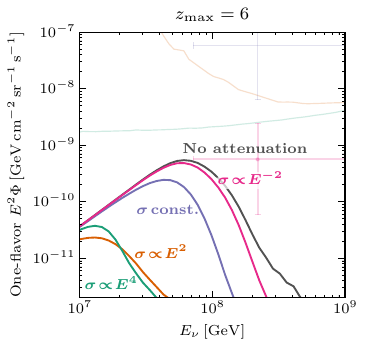}
    \caption{Diffuse neutrino flux for different energy-dependent cross sections over DM mass: $\sigma/M_\mathrm{DM} = 10^{-23} \mathrm{cm^2/GeV} \times (E_\nu^0/ 10^8\,\mathrm{GeV})^n$, assuming a particular model of UHE neutrino production and for a maximum redshift of cosmic ray sources $z_{\rm max}=4$ (left plot) or (right plot). See text for details. }
 \label{fig:UHEnu-attenuation-spectrum}
\end{figure}

Assuming that the event KM3-230213A is of diffuse origin, and inferring the corresponding flux by combining with the upper limits from IceCube and the Pierre Auger Observatory, one can derive an upper limit on the scattering cross section over the mass. The effect of the attenuation on the flux is illustrated in Fig.~\ref{fig:attenuation-KM3-event}, for an energy-independent cross section and considering an NFW profile. The line labeled ``No attenuation'' corresponds to the ``Joint'' fit shown in Fig.~\ref{fig:landscape}. As the cross section increases, the necessary injected flux to explain the detection of the event KM3-230213A increases, and it would eventually violate the Waxman-Bahcall bound. Namely, the state-of-the-art mechanisms of neutrino production in astrophysical sources are incapable of producing the event KM3-230213A, since even under the most aggressive assumptions for production, the DM-$\nu$ scatterings would attenuate the flux to unobservable levels. The upper limit reads $\sigma/M_{\rm DM}\lesssim 4\times 10^{-22}\,{\rm cm}^2/{\rm GeV}$. It is important to note that this limit is fairly robust against astrophysical assumptions, since the flux attenuation depends exponentially on $\sigma/M_{\rm DM}$. This is also illustrated in Fig.~\ref{fig:UHEnu-attenuation-spectrum}: if $\sigma/M_\mathrm{DM}$ is larger than our limit by a factor $\sim 2$, the unattenuated flux should violate the Waxman-Bahcall bound by about two orders of magnitude. On top of that, we conservatively include only attenuation by the Milky Way.

\begin{figure}[hbtp]
    \centering
    \includegraphics[width=0.9\linewidth]{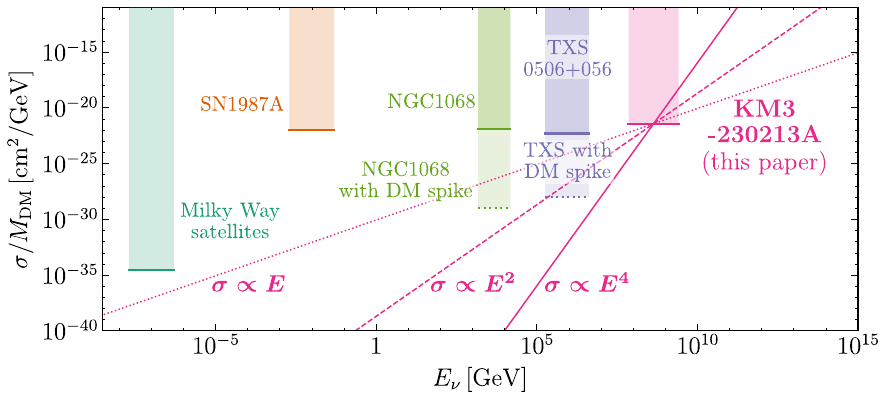}
    \caption{Limits on DM-$\nu$ scattering from the KM3-230213A event and from previous works. We assume that  KM3-230213A  has a diffuse origin, although the results are similar for a point source (under different assumptions, see text). We also show the extrapolation of the limits to other energies assuming simple energy dependence of the cross section.}
    \label{fig:compilation}
\end{figure}

We show in Fig.~\ref{fig:compilation} a compilation of the limits on the scattering cross section over DM mass as a function of the neutrino energy. At the highest energies, we present the limit derived in this work from the assumption that KM3-230213A is of diffuse origin (similar limits can be obtained if the event originates from a point source \textit{and} exponentially large attenuation is \textit{a priori} excluded). At lower energies, we show the limits from the blazar TXS 0506+056, excluding~\cite{Choi:2019ixb,Kelly:2018tyg,Alvey:2019jzx} or including~\cite{Ferrer:2022kei,Cline:2022qld} the attenuation from a DM spike surrounding the supermassive black hole at its center, the limits from the Seyfert Galaxy NGC 1068~\cite{Cline:2023tkp}, the limits from  from supernova neutrinos \cite{Raffelt:1996wa,Heston:2024ljf} and finally the limits from Milky Way satellite galaxies~\cite{Crumrine:2024sdn}. We also show for reference the scaling of the limits when the cross section has a simple power-law dependence with neutrino energy. Clearly, if this dependence holds over several decades of energy, the detection of UHE neutrinos poses very stringent constraints on DM-neutrino interactions at lower energies.

In many models, one expects indeed a non-trivial dependence of the cross section with energy (see Ref.~\cite{Arguelles:2017atb} for explicit examples, and Ref.~\cite{Dev:2025tdv} for a critical overview of models with observable cross sections). For example, the scattering cross section with a fermionic target at rest mediated by a scalar particle with mass $m_\phi$ grows as $E_\nu$ when $E_\nu\ll m_\phi$ if the target mass is much smaller than $E_\nu$, or as $E_\nu^2$ if the target mass is much higher than $E_\nu$. It then reaches a plateau when the center-of-mass energy of the collision is comparable to $m_\phi$, and it finally decreases with energy due to unitarity. As another example, the cross section with a scalar target mediated by a heavy fermion grows as $E_\nu^3$ if the target mass is much smaller than $E_\nu$, or as $E_\nu^4$ if it is much larger. Further, the cross section may display resonances, akin to the Glashow resonance when a very energetic electron antineutrino collides with an electron at rest reaching a center-of-mass energy comparable to the $W$ mass \cite{Glashow:1960zz}. Thus, depending on the particular model, UHE neutrino attenuation may be the leading probe of DM-$\nu$ interactions.

Further, future experiments should detect many more UHE neutrinos, which will also allow to determine their energy spectrum. The impact of the flux attenuation due to DM scatterings is shown in Fig.~\ref{fig:UHEnu-attenuation-spectrum}. We illustrate the effect with an unattenuated neutrino flux obtained using the public code \texttt{CRPropa} \cite{AlvesBatista:2016vpy}, for the 2SC-uhecr model presented in \cite{Ehlert:2023btz}, and which was used in Ref.~\cite{KM3NeT:2025vut} as a benchmark to explain the origin of  KM3-230213A. It assumes the injection of cosmic-ray protons by sources with luminosity $0.12\times 10^{44}\,{\rm  erg}/{\rm Mpc}^3/{\rm year}$, a spectrum following a power law proportional to $E^{-0.25}$ in the energy range  5-10 EeV, and a source distribution as a function of redshift proportional to $(1+z)^5$ up to a maximum redshift $z_{\rm max}=4$ (left plot) or 6 (right plot). The impact of the attenuation and its energy dependence is apparent from the plot: the peak of the distribution is shifted to lower energies, and the cut-off energy also decreases. The energy dependence of the cross section is also relevant, as an enhanced cross section at high energies suppresses the flux of neutrinos produced at high redshift, which for the same present-day energy had higher energies at production.

This plot illustrates that very aggressive astrophysical assumptions in the form of an assumed cosmic-ray spectrum, chemical composition, and redshift-dependent injection; would allow to tightly probe neutrino-DM interactions. More importantly, it highlights the relevance of constraining neutrino-DM interactions, as one of the main astrophysics goals of UHE neutrino detection is to learn about the sources, chemical composition, and redshift evolution of UHE cosmic rays. As the figure shows, this goal can be hindered by neutrino-DM interactions.

\section{Complementarity of UHE neutrino detectors}
\label{sec:complementarity}

In Section \ref{sec:attenuation_general} above, we have discussed the anisotropy in the neutrino flux at the surface of the Earth due to the attenuation by scattering with DM in the Milky Way halo. Testing this anisotropy would be a very robust probe of neutrino-DM scattering. This feature has been exploited with high-energy astrophysical neutrino data~\cite{IceCube:2022clp, Arguelles:2017atb} but, as we discuss below, requires a different approach at UHE energies.

Before reaching the detector, UHE neutrinos must traverse the Earth. This attenuates the neutrino flux by interactions with the nuclei at Earth. Thus, different experiments are generically sensitive to different fractions of the sky. The combination of different experiments at different locations is hence mandatory to look for anisotropies induced by DM-$\nu$ scattering.

\begin{figure}[hbtp]
    \centering
    \includegraphics[width=0.5\linewidth]{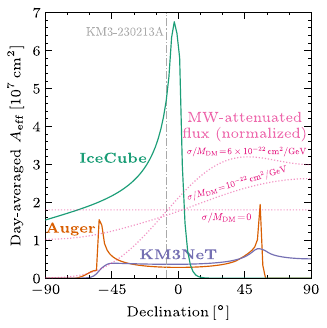}
    \caption{Day-averaged effective area of IceCube, the Pierre Auger Observatory and KM3NeT, as a function of the declination of the arriving neutrino. We also show as comparison the expected angular dependence of a diffuse UHE neutrino flux for different scattering cross sections over DM mass, due to attenuation in the Milky Way halo and assuming an NFW profile; together with the incoming direction of the KM3-230213A event. }
    \label{fig:eff_areas}   
\end{figure}

With current and several future detectors, the UHE neutrino sensitivity is maximized near the horizon. This enlarges the amount of matter traversed by neutrinos, increasing the probability for them to produce a detectable interaction, while avoiding Earth attenuation at angles well below the horizon. Therefore, experiments ``sweeping'' the sky as the horizon moves against the fixed stars are well-suited to be sensitive at different directions in the sky. We show in Fig.~\ref{fig:eff_areas} the day-averaged effective areas ({\it i.e.}, integrating over the right ascension) of IceCube, the Pierre Auger Observatory and KM3NeT as a function of the declination. KM3NeT (located at a latitude of $36.27^\circ$ N) has sensitivity to a wide range of declinations, being only blind to UHE neutrinos arriving with $-90^\circ\leq \delta\lesssim 36^\circ$, since this region of the sky remains below the horizon over the whole day. IceCube (located at a latitude of $90^\circ$ S) is, in contrast, practically blind to neutrinos arriving from positive declinations, and presents a high peak of sensitivity at $\delta\simeq 0$, since this direction lies at the horizon of the South Pole the whole day. For reference, in the figure we show the incoming direction of the KM3-230213A event, underscoring the tension of the observation of this event with the constraints from IceCube. Finally, the Pierre Auger Observatory (located at a latitude of $35.0^\circ$ S) is only sensitive to neutrinos from a very narrow range of incoming directions~\cite{PierreAuger:2019azx}. Although the effect is partly alleviated by the rotation of the Earth, the day-averaged effective area is still rather angle-dependent. Notice that the effective area of KM3NeT is a factor of $\sim 10$ smaller than the one from IceCube, even though they are both km$^3$ experiments, the reason being that KM3NeT is still under construction. 

\begin{figure}[hbtp]
    \centering
    \includegraphics[width=0.5\linewidth]{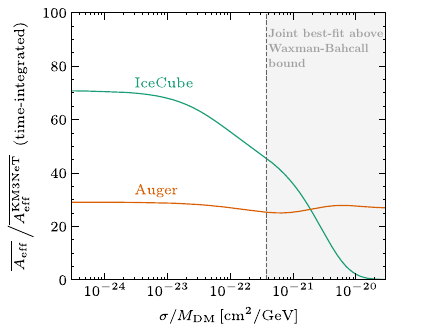}
    \caption{Effective sensitivity of IceCube and the Pierre Auger Observatory, relative to KM3NeT, of a diffuse UHE neutrino flux as a function of the scattering cross section over DM mass, due to the interplay of the effective area and arrival direction of the neutrino flux attenuated by DM scatterings in the MW halo assuming a NFW profile. }
    \label{fig:eff_areas-ratios}
\end{figure}

The figure also shows the expected astrophysical UHE neutrino flux for different values of the scattering cross section over DM mass. Clearly, for an unattenuated flux (purely isotropic) IceCube has a very good sensitivity. In turn, if the flux is attenuated by scattering with DM in the Milky Way halo, the flux at negative declinations decreases (see Fig.~\ref{fig:MW-map} for the angular dependence of the DM column density), and IceCube loses sensitivity. Other instruments, such as KM3NeT or Auger, that are sensitive to UHE neutrinos coming from positive declinations, may observe signals while escaping detection at IceCube. 

This is further illustrated in Fig.~\ref{fig:eff_areas-ratios}, which shows the ratios among the effective areas of IceCube and KM3NeT (green), and Auger and KM3NeT (orange) as a function of $\sigma/M_{\rm DM}$ and integrated over the exposure times of each detector.  The relative exposures of Auger and KM3NeT are roughly independent of $\sigma/M_{\rm DM}$, due to their locations on Earth. In turn, for small cross sections IceCube has a much bigger effective area than KM3NeT, but this decreases for large cross sections because IceCube is most sensitive to the region in the sky with the largest attenuation. This interplay of the effective areas of both experiments could in principle open the intriguing possibility that KM3NeT could observe UHE neutrinos, while escaping detection at IceCube, and thereby potentially explain the observation of KM3‑230213A. Quantitatively, however, the difference in effective areas without attenuation is so large that the impact of DM attenuation is mild: large suppressions for IceCube would require fluxes exponentially violating the Waxman-Bahcall bound, plus Auger has significant sensitivity in a sky region with little attenuation (see Figs.~\ref{fig:eff_areas} and \ref{fig:eff_areas-ratios}). Using a saturated Poisson likelihood test~\cite{Baker:1983tu}, we find that the tension between all three experiments can reduce from $\sim 3\sigma$~\cite{Li:2025tqf, KM3NeT:2025ccp, IceCubeCollaborationSS:2025jbi} to, at most, $\sim 2.8\sigma$. Nevertheless, this is an important effect to keep in mind should further observations of UHE neutrinos with comparable detectors be in mild apparent tension among them.

\section{Conclusions}
\label{sec:conclusions}

We have presented a comprehensive analysis of the attenuation of a diffuse neutrino flux  due to scattering with DM particles in the intergalactic medium and in the Milky Way halo. We have argued that when the scattering cross section over DM mass is sizable, the attenuation in the Milky Way generates a suppressed neutrino flux with a characteristic anisotropy.

We have then studied the implications of DM-$\nu$ interactions for the flux, energy spectrum and arrival direction of ultra-high energy neutrinos, assuming that they are of diffuse origin. Concretely, we have shown that the Waxman-Bahcall limit becomes more stringent in the presence of DM-$\nu$ interactions. This implies that, with mild astrophysical assumptions, limits on the DM-$\nu$ interaction cross section can be placed even if the unattenuated flux is unknown. Further, we have investigated the implications of the detection of the event KM3‑230213A for the DM-$\nu$ interactions, assuming that the event is of diffuse origin. This allows to exclude DM-neutrino interaction cross sections $\sigma/M_\mathrm{DM} \gtrsim 4 \times 10^{-22} \, \mathrm{cm^2/GeV}$ at unprecedentedly high energies. This limit is obtained assuming an NFW DM profile; other profiles would lead to comparable limits. If the event is indeed of diffuse origin, future instruments should detect hundreds of UHE neutrinos, which will allow to perform a detailed spectral and  morphological analysis of the signal, that encode invaluable information about the magnitude and energy dependence of the scattering cross section. Finally, we have highlighted the complementarity of different neutrino detectors at different latitudes in the Earth, in order to probe DM-neutrino interactions  through the anisotropy in the arrival directions of UHE neutrinos due to their passage through the DM halo of our Galaxy and the Earth.

\section*{Note added}
During the last stages of this work, we became aware of the preprints \cite{Bertolez-Martinez:2025trs,Mondol:2025uuw} which partly overlap with our analysis. Although their main focus is on the KM3NeT event and a point-source origin of it, when comparable, our conclusions agree.

\section*{Acknowledgments}
The work of AI is supported by the Collaborative Research Center SFB1258 and by the Deutsche Forschungsgemeinschaft (DFG, German Research Foundation) under Germany's Excellence Strategy - EXC-2094 - 390783311. The work of IE is supported by the Spanish MCIN/AEI/10.13039/501100011033 grants PID2021-123703NB-C21 and PID2022-136510NB-C33, and by the Basque Government grant IT1628-22. IE would like to thank the Technical University of Munich for hospitality during the first stages of this work.

\appendix
\renewcommand\thefigure{\thesection\arabic{figure}}
\renewcommand\theequation{\thesection\arabic{equation}}

\setcounter{figure}{0}
\setcounter{equation}{0}

\section{Evaluation of the effective areas}
\label{app:Aeff}

In this Appendix, we detail how we compute the effective areas for the relevant experiments discussed in the main text. While for Auger the effective area as a function of neutrino energy and incoming direction is publicly available~\cite{PierreAuger:2019azx}, to our understanding for IceCube and KM3NeT only sky-averaged effective areas exist. Directional effects are particularly relevant for UHE neutrino attenuation by DM, so a proper implementation of angle-dependent effects is mandatory.

We compute the angular dependence of the IceCube and KM3NeT effective areas following Ref.~\cite{Gaisser:2016uoy} (see Ref.~\cite{Palmisano:2025abd} for a more detailed approach using a transport equation). The effective area for a neutrino with energy $E_\nu$ and zenith angle $\theta$ that interacts inside the detector is given by
\begin{equation}
    A_\mathrm{eff}^\mathrm{in}(E_\nu, \theta) = \sigma(E_\nu) N_\mathrm{nuc}\varepsilon(E_\nu, \theta) e^{-\sigma(E_\nu) \int n(s, \theta) \, \mathrm{d}s} \, ,
\end{equation}
with $\sigma$ the neutrino-nucleon interaction cross section, $N_\mathrm{nuc}$ the number of nucleons in the detector, $\varepsilon$ the detection efficiency, and $n(s, \theta)$ the number density of nucleons along the traversed Earth path. The last term takes into account neutrino attenuation by Earth. In all our calculations, we assume the Preliminary Earth Reference Model (PREM)~\cite{Dziewonski:1981xy}. 

In turn, for a neutrino that interacts outside the detector and produces a throughgoing lepton ($\mu$ or $\tau$) that traverses the detector,
\begin{equation}
    A_\mathrm{eff}^\mathrm{through}(E_\nu, \theta) = A(\theta) P_\ell(E_\nu)\varepsilon(E_\nu, \theta) e^{-\sigma(E_\nu) \int n(s, \theta) \, \mathrm{d}s} \, ,
\end{equation}
with $A(\theta)$ the projected area of the detector perpendicular to the neutrino direction, and $P_\ell(E_\nu, \theta)$ the probability for a neutrino with energy $E_\nu$ to produce a throughgoing lepton that reaches the detector. The total effective area is given by
\begin{equation}
    A_\mathrm{eff}(E_\nu, \theta) = A_\mathrm{eff}^\mathrm{in}(E_\nu, \theta) + A_\mathrm{eff}^\mathrm{through}(E_\nu, \theta)\, .
\end{equation}
To compute $A(\theta)$, we approximate the IceCube and KM3NeT detectors as cylinders with volumes of $1\,\mathrm{km^3}$ and $0.15\,\mathrm{km^3}$, and heights of $1\,\mathrm{km}$ and $0.7\,\mathrm{km^3}$; respectively~\cite{KM3NeT:2025npi}.

For muons, $P_\mu(E_\nu,  \theta)$ is given by~\cite{Gaisser:2016uoy}
\begin{equation}
    P_\mu(E_\nu,  \theta) = \int_{E_\mathrm{th}}^\infty \mathrm{d}E_\mu \, \frac{N_A}{\alpha(1+E_\mu/\epsilon)}\int_{y'_\mathrm{min}}^{y'_\mathrm{max}}\mathrm{d}y' \, \frac{\mathrm{d}\sigma}{\mathrm{d}y'} \, ,
\end{equation}
where $E_\mu$ is the energy of the throughgoing muon, $N_A$ is Avogadro's number (equal to the number of nucleons per gram of matter), $\alpha = 2\times 10^{-3}\,\mathrm{GeV/(g/cm^2)}$ is the muon ionization energy loss rate, $\epsilon=440\,\mathrm{GeV}$ is the muon energy where radiative energy losses equal ionization energy losses~\cite{ParticleDataGroup:2024cfk}, $E_\mathrm{th}$ is the smallest detectable muon energy (its effect is negligible as long as $E_\mathrm{th}\ll E_\nu,\,\epsilon$; which we assume to hold), $y' \equiv 1 - E_\mu'/E_\nu$ with $E_\mu'$ the muon energy before energy losses, and $\sigma$ is the neutrino-nucleon interaction cross section. $y'_\mathrm{max} = 1 - E_\mu/E_\nu$, and 
\begin{equation}
    y'_\mathrm{min} = \max\left[0,\, 1 - \frac{(E_\mu + \epsilon)e^{X_\mathrm{max}(\theta)/\xi} - \epsilon}{E_\nu}\right] \, ,
\end{equation}
where $\xi \equiv \epsilon/\alpha$ and $X_\mathrm{max}(\theta) = \int_0^\infty \rho(s, \theta) \, \mathrm{d}s$ is the integrated grammage along the incoming direction, with $\rho$ mass density. This expression can be derived by following Ref.~\cite{Gaisser:2016uoy} but noting that the maximum integrated grammage along the incoming direction is necessarily finite.

For $\tau$ leptons, $\tau$ decay has to be taken into account, leading to 
\begin{equation}
    P_\tau(E_\nu,  \theta) = \int_{E_\mathrm{th}}^\infty \mathrm{d}E_\tau \, \frac{N_A}{\alpha(1+E_\tau/\epsilon)} e^{-\frac{m_\tau}{\rho \lambda_\tau} \frac{\xi}{\epsilon} \ln \left[ \frac{E_\tau' (E_\tau + \epsilon)}{E_\tau (E_\tau' + \epsilon)}\right]}\int_{y'_\mathrm{min}}^{y'_\mathrm{max}}\mathrm{d}y' \, \frac{\mathrm{d}\sigma}{\mathrm{d}y'} \, ,
\end{equation}
where for $\tau$ leptons $\epsilon = 3.4\times10^3\,\mathrm{GeV}$~\cite{Dutta:2005yt}, $m_\tau$ is the $\tau$ mass, and $\lambda_\tau \simeq 8.7 \times 10^{-3}\,\mathrm{cm}$ is the $\tau$ decay length. This expression can be derived by following Ref.~\cite{Gaisser:2016uoy} but noting that the $\tau$ lepton flux gets attenuated after travelling a distance $s$ by a factor $e^{-\frac{m_\tau}{\lambda_\tau} \int_0^s \frac{1}{E(s)}\, \mathrm{d}s}$, and changing variables from $s$ to integrated grammage $X$.

We have checked that following this procedure gives a very good approximation to the angular dependence of lower-energy publicly available IceCube effective areas~\cite{IceCube:2019cia}. 

The procedure described above allows to compute the effective area up to the detection efficiency $\varepsilon(E_\nu, \theta)$. To estimate it, we assume that, at the high energies relevant for UHE neutrino detection, the angular dependence of $\varepsilon$ is negligible compared to that of Earth absorption and throughgoing leptons. We thus obtain $\varepsilon(E_\nu)$ by requiring the sky-averaged effective areas to match those in Refs.~\cite{IceCubeCollaborationSS:2025jbi, KM3NeT:2025npi}.

\bibliographystyle{JHEP}
\bibliography{References}
\end{document}